%% file: main.tex
\documentclass[conference]{IEEEtran}

\input{./sections/preamble}
\usepackage{fst}
\graphicspath{{figures/}}

\begin{document}

\input{./sections/acronyms}

\title{Protograph-Based LDPC Code Design for Probabilistic Shaping with On-Off Keying}
\author{\IEEEauthorblockN{Alexandru Dominic Git$^{\dagger}$, Bal\'azs Matuz$^{\dagger}$, Fabian Steiner$^{\ddagger}$}
\IEEEauthorblockA{$^\dagger$Institute of Communications and Navigation, German Aerospace Center (DLR), Germany\\$^\ddagger$Institute for Communications Engineering, Technical University of Munich, Germany\\Email: a.git@tum.de, balazs.matuz@dlr.de, fabian.steiner@tum.de}}
\maketitle

\begin{abstract}
This work investigates protograph-based \ac{LDPC} codes for the \ac{AWGN} channel with \ac{OOK} modulation. A non-uniform distribution of the \ac{OOK} modulation symbols is considered to improve the power efficiency especially for low \acp{SNR}. To this end, a specific transmitter architecture based on time sharing is proposed that allows probabilistic shaping of (some) \ac{OOK} modulation symbols. Tailored protograph-based \ac{LDPC} code designs outperform standard schemes with uniform signaling and off-the-shelf codes by \SI{1.1}{dB} for a transmission rate of \SI{0.25}{bits/channel use}.
\end{abstract}

\acresetall

\input{./sections/introduction}
\input{./sections/model}

\input{./sections/design}
\input{./sections/results}
\input{./sections/conclusions}

\appendices
\input{./sections/appendix}


\end{document}

%% file: sections/preamble.tex
\usepackage{amsfonts,amsbsy}
\usepackage[nolist]{acronym}
\usepackage{graphicx,cite,amssymb,amsmath}
\usepackage[caption=false]{subfig}
\usepackage{tabularx}
\usepackage{hhline}
\usepackage{dsfont}
\usepackage[dvipsnames]{xcolor}
\usepackage{mathrsfs,bm}
\usepackage{mathtools}  
\mathtoolsset{showonlyrefs=true}
\usepackage{tikz}
\usepackage{pgf}
\usepackage{pgfplots}
\pgfplotsset{compat=1.14}
\usetikzlibrary{matrix,positioning,shapes,arrows,decorations,calc,fit}
\usetikzlibrary{decorations.pathreplacing}
\usetikzlibrary{spy,backgrounds}
\usetikzlibrary{external}
\usetikzlibrary{patterns}
\usetikzlibrary{shapes,arrows}
\usetikzlibrary{shadows.blur}
\usetikzlibrary{shapes.symbols}
\usetikzlibrary{
	pgfplots.groupplots,
	matrix
}
\usetikzlibrary{intersections}

\usepackage{pgfplotstable}
\usepackage{adjustbox}
\usepackage{multirow}
\usepackage{units}
\usepackage{verbatim}

    \newcommand{\matB}{\bm{B}}

    \newcommand{\SNR}{E_s/N_0}
    
    \newcommand{\AS}{A_{\mathsf{S}}}
    \newcommand{\AU}{A_{\mathsf{U}}}

    \newcommand{\XS}{X_{\mathsf{S}}}
    \newcommand{\XU}{X_{\mathsf{U}}}

    \newcommand{\expe}[1]{\mathsf{E}\left[ #1 \right]}
    
    \newcommand{\entr}[1]{\mathsf{H}\left(#1\right)}
    \newcommand{\entrinv}[1]{\mathsf{H}^{-1}\left(#1\right)}

    \newcommand{\YS}{Y_{\mathsf{S}}}
    \newcommand{\YU}{Y_{\mathsf{U}}}

    \newcommand{\sigS}{\sigma^2_{ \mathsf S}}
    \newcommand{\sigU}{\sigma^2_{ \mathsf U}}
    \newcommand{\sigSsur}{\tilde \sigma^2_{ \mathsf S}}
    
    \newcommand{\capa}{\mathsf{C}}
    \newcommand{\mi}[2]{\mathsf{I}\left(#1;#2\right)}
    \newcommand{\RC}{\mathsf{R_{C}}}

    \newcommand{\RTX}{\mathsf{R_{TX}}}
    \newcommand{\RDM}{\mathsf{R_{DM}}}
    \newcommand{\RTS}{\mathsf{R_{TS}}}

    \newcommand{\pb}{\mathsf{p}_0}
    \newcommand{\pp}{\mathsf{p}_1}

\definecolor{TUMBlack}{cmyk}{0,0,0,1}     
\definecolor{TUMWhite}{cmyk}{0,0,0,0}     

\definecolor{TUMBlue} {cmyk}{1,0.43,0,0}  

\definecolor{TUMDarkBlue}   {cmyk}{1,0.57,0.12,0.5}      
\definecolor{TUMDarkerBlue} {cmyk}{1,0.54,0.04,0.19}     
\definecolor{TUMMediumBlue} {cmyk}{0.9,0.48,0,0}         
\definecolor{TUMLighterBlue}{cmyk}{0.65,0.19,0.01,0.04}  
\definecolor{TUMLightBlue}  {cmyk}{0.42,0.09,0,0}        

\definecolor{TUMDarkGray}  {cmyk}{0,0,0,0.8}  
\definecolor{TUMMediumGray}{cmyk}{0,0,0,0.5}  
\definecolor{TUMLightGray} {cmyk}{0,0,0,0.2}  

\definecolor{TUMGreen} {cmyk}{0.35,0,1,0.2}         
\definecolor{TUMOrange}{cmyk}{0,0.65,0.95,0}        
\definecolor{TUMIvory} {cmyk}{0.03,0.04,0.14,0.08}  

\definecolor{TUMBeamerYellow}    {rgb}{1.00,0.71,0.00}  
\definecolor{TUMBeamerOrange}    {rgb}{1.00,0.50,0.00}  
\definecolor{TUMBeamerRed}       {rgb}{0.90,0.20,0.09}  
\definecolor{TUMBeamerDarkRed}   {rgb}{0.79,0.13,0.25}  
\definecolor{TUMBeamerBlue}      {rgb}{0.00,0.60,1.00}  
\definecolor{TUMBeamerLightBlue} {rgb}{0.25,0.75,1.00}  
\definecolor{TUMBeamerGreen}     {rgb}{0.57,0.67,0.42}  
\definecolor{TUMBeamerLightGreen}{rgb}{0.71,0.79,0.51}  

%% file: sections/acronyms.tex
\begin{acronym}
    \acro{5G}{fifth generation of mobile networks}
	\acro{AMI}{average mutual information}
	\acro{AR4A}{accumulate-repeat-$4$-accumulate}
	\acro{ARQ}{automatic repeat-request}
	\acro{AWGN}{additive white Gaussian noise}
	\acro{BP}{belief propagation}
	\acro{BCH}{Bose-Ray-Chaudhuri-Hocquenghem}
	\acro{BCJR}{Bahl-Cocke-Jelinek-Raviv}
	\acro{BEC}{binary erasure channel}
	\acro{BEEC}{binary error and erasure channel}
	\acro{BER}{bit error rate}
	\acro{BICM}{bit-interleaved coded modulation}
	\acro{BPSK}{binary phase shift keying}
	\acro{BSC}{binary symmetric channel}
	\acro{CER}{codeword error rate}
	\acro{CCSDS}{Consultative Committee for Space Data Systems}
	\acro{CM}{coded modulation}
	\acro{CN}{check node}
	\acro{CRC}{cyclic redundancy check}
	\acro{DLR}{Deutsches Zentrum für Luft- und Raumfahrt}
	\acro{DVB-RCS}{Digital Video Broadcasting -- Return Channel via Satellite}
	\acro{DVB-SH}{Digital Video Broadcasting -- Satellite Services to Handheld Devices}
	\acro{DE}{density evolution}
	\acro{ESR}{erroneous second ratio}
	\acro{ESA}{European Space Agency}
	\acro{EVBD}{enhanced verification-based decoding}
	\acro{EXIT}{extrinsic information transfer}
	\acro{FHT}{fast Hadamard transform}
	\acro{G-LDPC}{generalized low-density parity-check}
	\acro{GE}{Gaussian elimination}
	\acro{GeIRA}{generalized irregular repeat-accumulate}
	\acro{GF}{Galois field}
	\acro{HT}{Hadamard transform}
	\acro{i.i.d.}{independent and identically distributed}
	\acro{IRA}{irregular repeat-accumulate}
	\acro{IS-95}{Interim Standard 95}
	\acro{IT}{iterative}
	\acro{LDPC}{low-density parity-check}
	\acro{LLR}{log-likelihood ratio}
	\acro{LMSC}{land mobile satellite channel}
	\acro{MAC}{media access control}
	\acro{MAP}{maximum a posteriori}
	\acro{MAW}{maximum accumulated weight}
	\acro{MBMS}{Multimedia Broadcast Multicast Service}
	\acro{MC}{maximum component}
	\acro{MCW}{maximum column weight}
	\acro{MDS}{maximum distance separable}
	\acro{ML}{maximum likelihood}
	\acro{MLC}{multilevel coding}
	\acro{ML-P}{maximum likelihood-pivoting}
	\acro{MP}{message passing}
	\acro{MPE-iFEC}{Multi-Protocol Encapsulation -- inter-burst Forward Error Correction}
	\acro{MR}{multiplicatively repeated}
	\acro{MTBL}{maximum tolerable burst length}
	\acro{NASA}{National Aeronautics and Space Administration}
	\acro{PEG}{progressive edge-growth}
	\acro{PHY}{physical layer}
	\acro{PMF}{probability mass function}
	\acro{PPM}{pulse-position modulation}
	\acro{qSC}{$q$-ary symmetric channel}
	\acro{RCB}{random coding bound}
	\acro{RCC}{recursive convolutional code}
	\acro{RM}{Reed-Muller}
	\acro{RS}{Reed-Solomon}
	\acro{RSC}{recursive systematic convolutional}
	\acro{r.v.}{random variable}
	\acro{SEME}{single-error multiple-erasures}
	\acro{SISO}{soft-input soft-output}
	\acro{SNR}{signal-to-noise ratio}
	\acro{SPB}{sphere packing bound}
	\acro{TUB}{truncated union bound}
	\acro{UMTS}{Universal Mobile Telecommunications System}
	\acro{VBD}{verification-based decoding}
	\acro{VN}{variable node}
	\acro{WE}{weight enumerator}
	\acro{WEF}{weight enumerator function}
	\acro{LASER}{light amplification by simulated emission of radiation}
	\acro{LDPC}{low-density parity-check}
	\acro{RS}{Reed-Solomon}
	\acro{EXIT}{extrinsic information transfer}
	\acro{PEXIT}{protograph extrinsic information transfer}
	\acro{LEO}{low-Earth orbit}
	\acro{GEO}{geostationary orbit}
	\acro{GTO}{geosynchronous transfer orbit}
	\acro{ISL}{inter-satelllite link}
	\acro{PPM}{pulse-position modulation}
	\acro{OOK}{on-off keying}
	\acro{ASK}{amplitude-shift keying}
	\acro{PSK}{phase-shift keying}
	\acro{APSK}{amplitude and phase-shift keying}
	\acro{QAM}{quadrature amplitude modulation}
	\acro{BPSK}{binary phase-shift-keying}
	\acro{QPSK}{quadrature phase-shift-keying}
	\acro{DLR}{German Aerospace Center}
	\acro{VN}{variable node}
	\acro{CN}{check node}
	\acro{CCDM}{constant-composition distribution matcher}
	\acro{DVB-S2}{Digital Video Broadcasting - Satellite 2}
	\acro{SNR}{signal-to-noise ratio}
	\acro{LLR}{log-likelihood ratio}
	\acro{PDF}{probabiliy distribution function}
	\acro{pdf}{probability density function}
	\acro{PMF}{probability mass function}
	\acro{BCH}{Bose-Chaudhuri-Hocquenghem}
	\acro{P-LDPC}{protograph-based low-density parity-check}
	\acro{DE}{differential evolution}
	\acro{MI}{mutual information}
	\acro{FEC}{forward error correction}
	\acro{APP}{a posteriori probability}
	\acro{FSO}{free-space optical}
	\acro{SE}{spectral efficiency}
	\acro{BER}{bit error rate}
	\acro{SPC}{single parity check}
	\acro{PS}{probabilistic shaping}
	\acro{PAS}{probabilistic amplitude shaping}
	\acro{CCSDS}{Consultative Committee for Space Data Systems}
	\acro{CRC}{cyclic redundancy check}
	\acro{LNT}{Institute for Communications Engineering} 
	\acro{RF}{radio frequency}
	\acro{i.i.d.}{independent identically distributed}
	\acro{RV}{random variable}
	\acro{TS}{time sharing}
	\acro{CM}{coded modulation}
	\acro{BICM}{bit-interleaved coded modulation}
	\acro{BMD}{bit-metric decoding}
	\acro{BRGC}{binary reflected Gray code}
	\acro{FER}{frame error rate}
	\acro{SMD}{symbol metric decoding}
	\acro{IM}{intensity modulation}
	\acro{SE}{spectral efficiency}
	\acro{NB-LDPC}{non-binary low-density parity-check}
	\acro{DM}{distribution matcher}
	\acro{DD}{direct detection}
\end{acronym}

%% file: sections/introduction.tex
\section{Introduction}\label{sec:Intro}

\acused{FSO}Free-space optical (FSO) communication has numerous advantages: large bandwidth, license free spectrum, high data rate, and easy deployment.
\acused{IM}Intensity modulation (IM) schemes, such as \ac{OOK} and \ac{PPM} are widely used for \ac{DD} receivers~\cite[Sec.~V]{khalighi_survey_2014}, since they do not require an optical phase-locked loop to track the carrier phase at the receiver. Non-coherent schemes are currently considered for deep-space communications, near earth communications and space-to-ground communications~\cite{MH05,Hemmati2011,CCSDS18}.

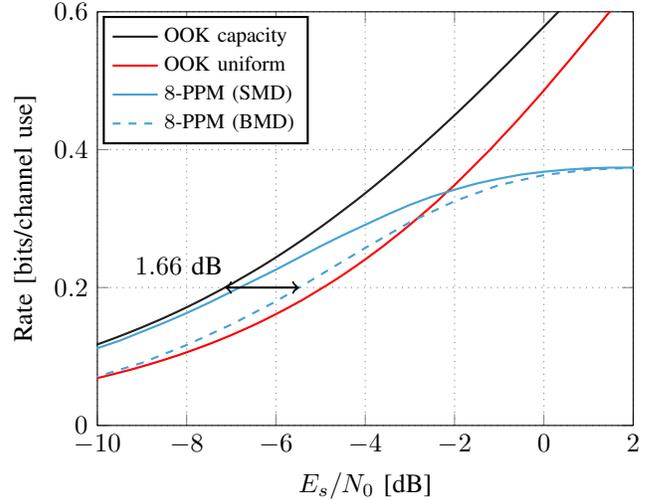
\begin{figure}[t]    
    \centering
    \footnotesize
    \tikzsetnextfilename{ook_rates_cmp}
    \input{./figures/ook_rates_cmp}
    \caption{Achievable rates for OOK and PPM.}
    \label{fig:ook_ppm}
\end{figure}

We study average power constrained \ac{AWGN} channels with low transmission power. For such channels, \ac{OOK} with a uniform distribution on the two levels shows a significant loss compared to optimal signaling using a non-uniform distribution. 
We can generate a non-uniform distribution by using \ac{PPM}. However, \ac{PPM} requires \ac{SMD} for good performance, i.e., the \ac{FEC} decoder must operate on the whole \ac{PPM} symbol. Instead, if binary codes with \ac{BMD} are considered, bit-wise soft-information is obtained by marginalizing over the bit-levels of the \ac{PPM} symbols and their correlation is not exploited, which generally leads to a performance loss. While this loss is small for some modulation schemes with a proper choice of the binary labeling (e.g., \ac{QAM} with  a \ac{BRGC}~\cite{gray1953pulse}), \ac{BMD} with \ac{PPM} experiences significant losses with respect to channel capacity~\cite{Herzog1997}. This is illustrated in Fig.~\ref{fig:ook_ppm}, where a gap of  almost \SI{1.7}{dB} between \ac{OOK} with a capacity achieving input distribution and $8$-\ac{PPM} with \ac{BMD} at a rate of \SI{0.2}{bits per channel use} (bpcu) is visible. To reduce this gap, iterations between the decoder and the demodulator have been considered~\cite{Herzog1997} but this increases receiver complexity. 

In general, the combination of \ac{PS} with \ac{FEC} is challenging as conventional schemes (e.g., \cite[Sec.~6.2]{gallager1968}, \cite{forney_trellis_1992}) place the shaping operation after \ac{FEC} encoding so that it needs to be reversed before (or performed jointly with) the \ac{FEC} decoding. This is prone to error propagation, synchronization issues~\cite[Sec.~IV-A]{forney_efficient_1984} or requires a joint and thus inflexible shaping/\ac{FEC} code design. Recently, \ac{PAS} was proposed, which avoids these difficulties by \emph{reverse concatenation}~\cite{w._g._bliss_circuitry_1981}. \ac{PAS} exploits the symmetry of the capacity achieving input distribution, but \ac{PAS} cannot be used for \ac{OOK}, as the optimal input distribution is not symmetric around the origin. Following the idea of \emph{sparse-dense} transmission~\cite{ratzer2013error}, we propose a \ac{TS} scheme which combines non-uniform signaling for \ac{OOK} with \ac{FEC}. Using a binary \ac{FEC} code of block length $n$ and code rate $\RC$ with systematic encoding, a number of $\RC\cdot n$ \ac{OOK} symbols is transmitted with a non-uniform distribution, while the remaining $(1-\RC)\cdot n$ parity bits are sent with a uniform distribution. A similar approach for coherent higher order modulations with non-binary \ac{LDPC} codes was suggested in~\cite{boutros_probabilistic_2017}, while binary codes were investigated for a nonlinear Fourier transform based, optical transmission system in~\cite{buchberger_probabilistic_2018} without a tailored \ac{LDPC} code design.

In this work, we describe a \ac{PS} approach for the average power constrained \ac{AWGN} channel with \ac{OOK} modulation via \ac{TS} and calculate achievable rates for this signaling strategy. We distinguish two cases. In the first one, both shaped and uniform symbols have the same amplitude. In the second case, the amplitudes may be chosen differently allowing an additional degree of freedom. For the \ac{P-LDPC} code design, we use \ac{EXIT} analysis with the surrogate approach of \cite{Steiner2016}. The proposed \ac{PS} scheme yields gains of up to \SI{1.1}{dB} with respect to uniform \ac{OOK} and off-the-shelf DVB-S2 codes~\cite{etsi2009dvb}.

%% file: figures/ook_rates_cmp.tex
\begin{tikzpicture}
\begin{axis}
[
width=0.48\textwidth,
height=0.3\textheight,
grid=both,
ymax = .6,
ymin = 0,
xmin = -10,
xmax =2,
ylabel={Rate [bits/channel use]},
xlabel={$\SNR$ [dB]},
grid = major,
legend cell align=left,
legend style={font=\footnotesize},
legend columns=1,
major grid style={dotted,gray},
legend style={at={(0.01,0.99)},anchor=north west,thick,font=\footnotesize},
]
\normalsize




\addplot[name path global=ook_shaped,color = TUMBlack, mark=none, mark options={solid}, thick]
table[x=Es/N0,y=se2,col sep=space,trim cells=true]  {./data/cap_uni_opt.txt};
\addlegendentry{OOK capacity};


\addplot[name path global=ook_uniform,color = red, mark=none, mark options={solid}, thick]
table[x=Es/N0,y=se1,col sep=space,trim cells=true]  {./data/cap_uni_opt.txt};
\addlegendentry{OOK uniform};

\addplot[name path global=ppm_smd, color = TUMLighterBlue, mark=none, mark options={solid}, thick]
table[x=SNR,y=C,col sep=space,trim cells=true] {./data/capacity_Hadamard8.txt};
\addlegendentry{$8$-PPM (SMD)};

\addplot[name path global=ppm_bmd,color = TUMLighterBlue,dashed, mark=none, mark options={solid}, thick]
table[x=SNR,y=C,col sep=space,trim cells=true] {./data/capacity_Hadamard8_BICM.txt};
\addlegendentry{$8$-PPM (BMD)};

%
%
%

\path[name path global=line] (-10,0.2) -- (2,0.2);
\path[name intersections={of=line and ook_shaped, name=p1}, name intersections={of=line and ppm_bmd, name=p2}];

\draw[<->,thick] let \p1=(p1-1), \p2=(p2-1) in (p1-1) -- (p2-1) node [below,xshift=-1.6cm,yshift=.55cm] {%
        \pgfplotsconvertunittocoordinate{x}{\x1}%
        \pgfplotscoordmath{x}{datascaletrafo inverse to fixed}{\pgfmathresult}%
        \edef\valueA{\pgfmathresult}%
        \pgfplotsconvertunittocoordinate{x}{\x2}%
        \pgfplotscoordmath{x}{datascaletrafo inverse to fixed}{\pgfmathresult}%
        \pgfmathparse{\pgfmathresult - \valueA}%
        \pgfmathprintnumber{\pgfmathresult} dB
};

\end{axis}
\end{tikzpicture}

%% file: sections/model.tex
\section{System Model and Optimal Signaling for OOK}

Consider transmission over an average power constrained \ac{AWGN} channel with
\begin{align}
Y = X + N
\end{align}
for $n$ channel uses. The Gaussian noise $N$ has zero mean and variance
$\sigma^2$. The \ac{OOK} constellation symbols $X$ are from the binary set $\mathcal{X} = \{0, A\}$.
The average power constraint is $\expe{X^2} \leq P$, where
\begin{equation}
\expe{X^2} =  A^2 P_X(A)\label{eq:power_constraint_general}.
\end{equation}
Without loss of generality, let $P = 1$. We define the
\ac{SNR} as $\SNR = 1/(2 \sigma^2)$.

An achievable rate is given by the mutual information $\mi{X}{Y}$ and the 
maximum achievable rate is the solution to the following optimization
problem
\begin{align} \label{eq:ook_capacity}
  \capa = \max_{P_X} \mi{X}{Y} \quad \text{subject to} \quad  A^2 P_X(A) \leq 1.
\end{align}
We refer to \eqref{eq:ook_capacity} as the ``OOK capacity'', which is shown in Fig.~\ref{fig:ook_ppm}. Note that the inequality constraint of the average power constraint is always active, so that the amplitude is $A=\sqrt{1/P_X(A)}$. If a uniform distribution is chosen, i.e., $P_X(0) = P_X(A) = \num{0.5}$, we observe a significant degradation in power efficiency.
\begin{figure*}[t] 
\begin{adjustbox}{max totalsize={1\textwidth}{1\textheight},center}
  \centering
    \tikzsetnextfilename{system_model}
    \input{./figures/system_model}
\end{adjustbox}
\caption{Block diagram of the proposed TS transceiver architecture for probabilistic shaping with OOK.} \label{fig:concatenation}
\end{figure*}
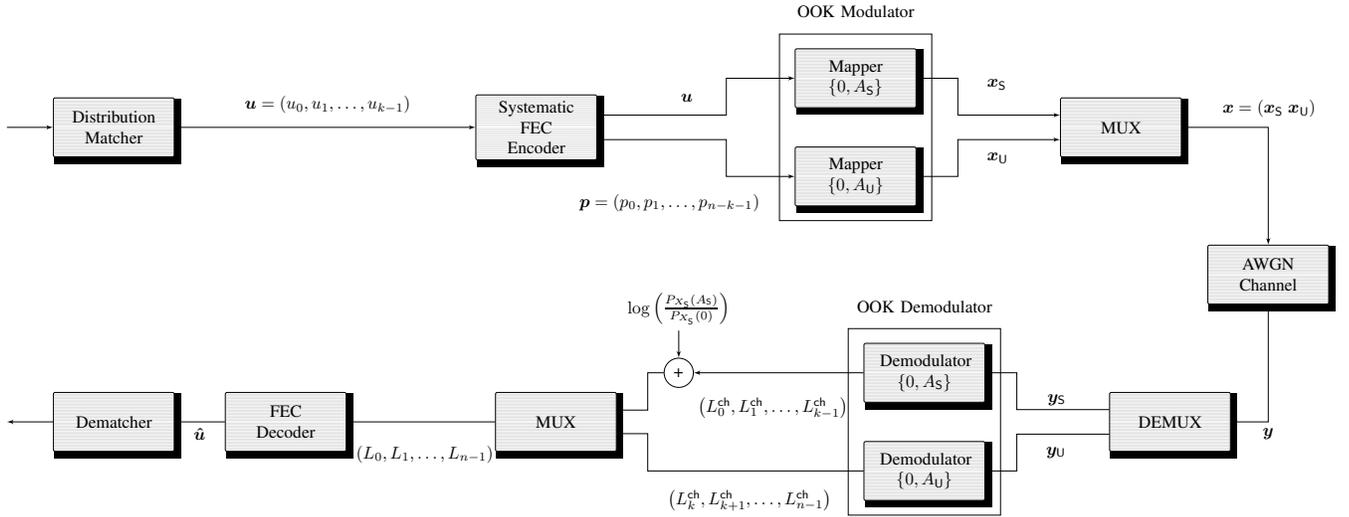

\section{Probabilistic Shaping via Time Sharing}
\label{sec:ps_via_ts}

We use a linear \ac{FEC} code of dimension $k$ and block length $n$. The
code rate is $\RC = k/n$. Its systematic generator matrix is of form $\bm{G} = (\bm{I}~\bm{P})$, where $\bm{I}$ is the $k\times k$ identity matrix and $\bm{P}$ is the $k\times (n-k)$ parity forming part. For encoding, the length $k$ information vector $\bm{u} \in \{0,1\}^k$ is multiplied with $\bm{G}$ yielding the codeword $\bm{c} = (\bm{u}~\bm{p})$ with $\bm{p} = \bm{u}\bm{P}$. 
The parity bits $\bm{p}$  are approximately uniformly distributed, since they are the result of a modulo-$2$ sum of
many information bits (see \cite[Sec.~IV-A]{Boecherer2015} for details). In contrast, the distribution of the information bits can be chosen at will, as explained later. This observation gives rise to a \ac{TS} sharing scheme which has been named \emph{sparse-dense} transmission in \cite{ratzer2013error,bocherer2012capacity}. 

In the following, we distinguish between a modulated information symbol $\XS$ and  modulated parity symbol $\XU$. We have $P_{\XS}=(\pb~\pp)$ and $P_{\XU}\approx(0.5~0.5)$. For the information part, i.e., for a number of $\RC n$ channel uses, we use the signaling set $\mathcal{X}_{\mathsf{S}} = \{0, A_{\mathsf{S}}\}$. For the remaining $(1-\RC)n$ channel uses involving the parity bits, the signaling set $\mathcal{X}_{\mathsf{U}} = \{0, A_{\mathsf{U}}\}$.

A \ac{CCDM} is used to realize the non-uniformly distributed symbols~\cite{Schulte2016}. The \ac{CCDM} encodes $k'$ uniformly distributed bits into
a length $k$ shaped information bit sequence $\bm{u}$ which is then \ac{FEC} encoded. The \ac{DM} is characterized by its matching rate
\begin{align}
\RDM = \frac{k'}{k}.\label{eq:Rdm}
\end{align}
For long $k$ the 
\ac{DM} rate \eqref{eq:Rdm} approaches the entropy of the output distribution \cite{Schulte2016}. Therefore, we may write $\RDM = \entr{\XS}$ for large $k$ and the overall transmission rate is 
\begin{align}
\RTX = \entr{\XS} \cdot \RC\label{eq:Rdm_asymptotic}.
\end{align}
Thus $\RTX$ is
directly related to $P_{\XS}(A_\mathsf{S}) = \pp$ via
\begin{align}
    \pp =\entrinv{\frac{\RTX}{\RC}}.\label{eq:ppulse}
\end{align}

For the general signaling set $\mathcal{X}$, the receiver performs soft-demapping and calculates the soft-information values
\begin{equation} \label{eq:llr}
L = \underbrace{\log \left( \frac{p_{Y|X}(y|A)}{p_{Y|X}(y|0)}\right)}_\text{{channel LLR}} + \underbrace{\log \left( \frac{P_X(A)}{P_X(0)}\right)}_\text{{prior}} .
\end{equation}
Note that the prior term is zero for the parity bits. The soft-information serves as an input to an \ac{LDPC} decoder which performs belief propagation decoding. The system setup is depicted in Fig.~\ref{fig:concatenation}.

\section{Rates for the Time Sharing Scheme}

\subsection{Transmission Rate}
\label{sec:transmission_rates}

An achievable rate of the \ac{TS} scheme is given by
\begin{align}
\RTS = \RC \mathsf{I}(\XS;\YS)  +(1-\RC)\mathsf{I}(\XU;\YU)\label{eq:achievable_rate_ts}.
\end{align}
From \eqref{eq:Rdm_asymptotic}, reliable communication is guaranteed as long as $\RTX \leq \RTS$. In the following, we distinguish two cases.

\subsection{Case 1: Same Pulse Amplitudes}
\label{sec:achievable_rates_ts1}

Consider the case where both pulse amplitudes are the same, i.e., $\AS=\AU=A$. The average power constraint \eqref{eq:power_constraint_general} is
\begin{align}
    \expe{\RC \XS^2+(1-\RC)\XU^2} = \left(\RC \pp + (1-\RC) \frac{1}{2}\right) A^2\label{eq:power_constraint_case1}
\end{align}
and the optimization problem for \eqref{eq:achievable_rate_ts} is
\begin{align}
\RTS_1^* = &\max_{\pp, A}\, \RTS \quad \\
&\text{subject to} \quad \left(\RC \pp + (1-\RC) \frac{1}{2}\right) A^2 \leq 1\label{eq:opt_ts1}.
\end{align}
As for \eqref{eq:ook_capacity}, the power constraint is always active. Thus, for a fixed $\pp$ we have $A=1/\sqrt{\RC \pp + (1-\RC)/2}$.

\subsection{Case 2: Individual Pulse Amplitudes}
\label{sec:achievable_rates_ts2}

We now permit different pulse amplitudes $\AS$ and $\AU$. The power constraint \eqref{eq:power_constraint_general} becomes
\begin{align}
    \expe{\RC \XS^2+(1-\RC)\XU^2} = \RC \pp \AS^2+ (1-\RC) \frac{1}{2}\AU^2 \label{eq:power_constraint_case2}.
\end{align}
Similar to the first case, we have
\begin{align}
\RTS_2^* = &\max_{\pp, A_{\mathsf{S}}, A_{\mathsf{U}}}\, \RTS \quad \\
    &\text{subject to}\quad  \RC \pp \AS^2+ (1-\RC) \frac{1}{2}\AU^2\leq 1 .\label{eq:opt_ts2}
\end{align}
Again, the average power constraint is always active. Thus, for a given $\pp$ the amplitude $\AU$ is given by $\AU=\sqrt{(1- \RC \pp \AS^2)/ ((1-\RC)/2)}$, where $\AS$ is subject to optimization.

\subsection{Numerical Comparison of Both Cases}
\label{sec:numerical_comparison_cases}

\begin{figure*}[t]
    \centering
    \footnotesize
    \subfloat[][$\RC=\num{0.5}$]{\tikzsetnextfilename{achievable_rates_ps-Rc=0.50}\input{./figures/achievable_rates_ps-Rc=0.50}}
 	\subfloat[][$\RC=\num{0.75}$]{\tikzsetnextfilename{achievable_rates_ps-Rc=0.75}\input{./figures/achievable_rates_ps-Rc=0.75}}
 	\caption{Achievable rates for the \ac{TS} scheme with different code rates $\RC$.}
 	\label{fig:TS}
 \end{figure*}
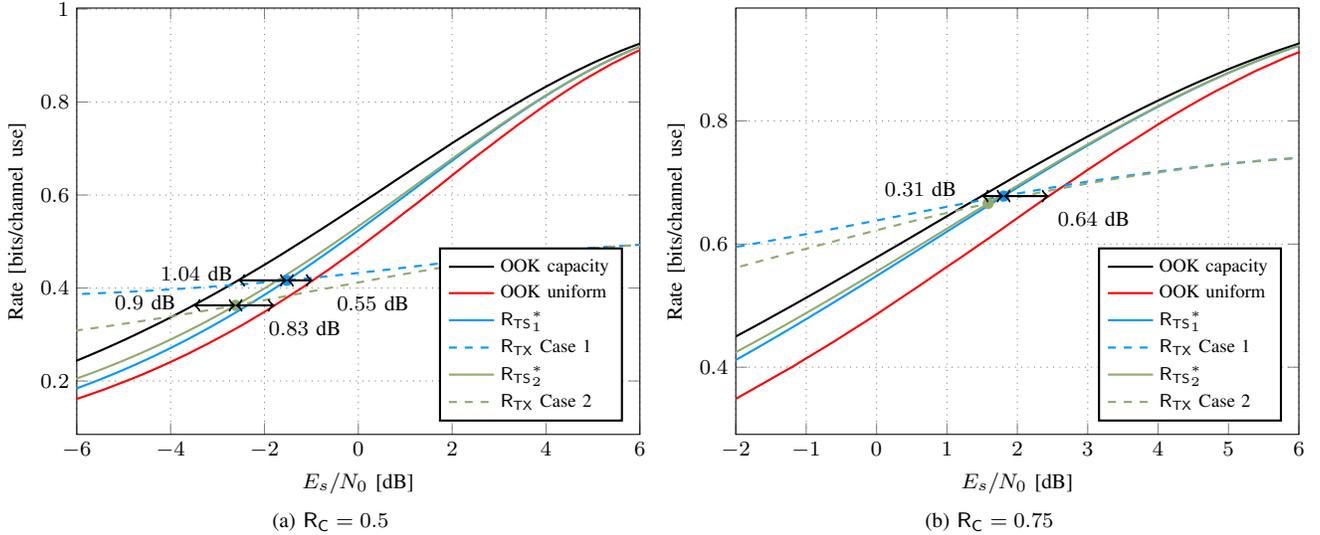

We plot the achievable rates for both time sharing schemes in Fig.~\ref{fig:TS} for the code rates $\RC = \num{0.5}$ and $\RC = \num{0.75}$. The dashed curves show the transmission rates \eqref{eq:Rdm_asymptotic} with the optimized pulse probability $\pp$ according to \eqref{eq:opt_ts1} and \eqref{eq:opt_ts2}. The crossing of the $\RTS$ and $\RTX$ curves indicates the optimal operating points for the chosen code rates.  Comparing \eqref{eq:ook_capacity} and \eqref{eq:achievable_rate_ts}, we observe that using a low code rate, i.e., $\RC=0.5$ in Fig.~\ref{fig:TS}, increases the gap as the fraction of transmission symbols with a uniform distribution also increases with lower $\RC$. The gap to the \ac{OOK} capacity is about \SI{1.0}{dB} for $\RC=\num{0.5}$, while it reduces to \SI{0.3}{dB} for $\RC = \num{0.75}$ code. These results motivate using a high rate code, even for low transmission rates. This requires using a pulse probability different from the optimal one from \eqref{eq:opt_ts1} or \eqref{eq:opt_ts2}. However, it provides an increased shaping gain due to the higher fraction of shaped symbols. For example, consider the first \ac{TS} scheme. In order to operate at $\RTX=\SI{0.25}{bpcu}$ as in Fig.~\ref{fig:TS}~(a),  instead of $\RC=0.5$ we may use $\RC=0.75$ with $\pp$ directly given by \eqref{eq:ppulse}. In the following, we discuss the choice of the code rate for a desired transmission rate.

\subsection{Signaling for a Fixed Transmission and \ac{FEC} Code Rate}
\label{sec:signaling_fixed_transmission_rate}

As pointed out in Sec.~\ref{sec:transmission_rates}, for a target transmission rate $\RTX$ and fixed code rate $\RC$, the probability $\pp$ is directly given by \eqref{eq:ppulse}. Thus, for the first \ac{TS} scheme, the average power constraint in~\eqref{eq:power_constraint_case1} determines $A$ and there are no additional degrees of freedom for the optimization in~\eqref{eq:opt_ts1}. The second \ac{TS} scheme has an additional degree of freedom by optimizing over either $\AS$ or $\AU$. 

A practical communication scheme usually uses of a family of channel codes of different rates.  For any target transmission rate we are interested in choosing the code rate to minimize the required $\SNR$. We proceed as follows.
\begin{enumerate}
    \item Consider a set $\mathcal{R}_\mathsf{C}$ of code rates.
    \item For a target $\RTX$, determine the $\SNR$ for all possible $\RC \in \mathcal{R}_\mathsf{C}$, such that $\RTX = \RTS_{i}^*, i \in \{1,2\}$. Since  $\RTX$ is fixed, for a certain $\RC$ the pulse probability $\pp$ is obtained from \eqref{eq:ppulse}.
    \item Among all $\RC \in \mathcal{R}_\mathsf{C}$ use the code rate $\RC^*$ that requires the smallest $\SNR$.
\end{enumerate}
As an example, consider the set of code rates $\mathcal{R}_\mathsf{C}=\{0.25, 0.33, 0.5, 0.67,0.75,0.8,0.9\}$. For different transmission rates in the range $\SI{0.2}{\bpcu}\leq \RTX \leq \SI{0.85}{\bpcu}$ we determine the required $\SNR$ for the code rates in $\mathcal{R}_\mathsf{C}$, and choose for each $\RTX$ the code rate $\RC^*$ with the lowest $\SNR$ requirement. Table~\ref{tab:RCstar} gives an overview of the code rates $\RC^*$ for some $\RTX$. The gray curves in  Fig.~\ref{fig:asymptotic_results} represent the corresponding achievable rates versus $\SNR$ for the first and second \ac{TS} schemes using code rates from Tab.~\ref{tab:RCstar}. Observe from the table that for the second \ac{TS} scheme it is beneficial to use high code rates, even if low transmission rates are targeted.
\begin{table}[t]
	\begin{center}
		\caption{Code rates $\RC^*$ for some $\RTX$.}
		\label{tab:RCstar}
		\begin{tabular}{ccc}
			\toprule
			\toprule
	             $\RTX$ & $\RC^*$ case $1$ & $\RC^*$ case $2$  \\
	       \midrule
	$0.2$& $0.33$  & $0.67$ \\	
	$0.25$& $0.5$  & $0.67$ \\	
	$0.33$& $ 0.5$ & $ 0.67$\\
    $0.5$& $0.67$  & $0.67$ \\
	$0.67$& $0.75$ & $ 0.8$ 	\\
	$0.75$& $0.8$  & $0.8$ 	\\
	$0.85$& $0.9$  & $0.9$ 	\\
	\bottomrule
		\bottomrule
		\end{tabular}
	\end{center}
\end{table}

%% file: figures/system_model.tex
\tikzset{
  -|-/.style={
    to path={
      (\tikztostart) -| ($(\tikztostart)!#1!(\tikztotarget)$) |- (\tikztotarget)
      \tikztonodes
    }
  },
  -|-/.default=0.5,
  |-|/.style={
    to path={
      (\tikztostart) |- ($(\tikztostart)!#1!(\tikztotarget)$) -| (\tikztotarget)
      \tikztonodes
    }
  },
  |-|/.default=0.5,
	ccNode/.style={rectangle,fill=white,fill opacity=.7, draw=black, very thick, minimum width=1.4cm, minimum height=1.4cm, text centered},
	interlNode/.style={rectangle,draw=black, very thick, inner sep=0.0em, minimum size=3.0em, text centered},
	sumNode/.style={circle,draw=black, very thick, inner sep=0.0em, minimum size=1.0em, text centered, node distance=0.25cm},
	conArrow/.style={->, ,rounded corners=3pt},
	con/.style={thick,rounded corners=3pt},
	inputNode/.style={circle, draw=black, thick, inner sep=0.0em, minimum size=2mm},
	conNode/.style={circle,draw=black, fill=black, minimum size=4pt, inner sep=0.0em},
	bnode/.style= {shade, blur shadow={shadow blur steps=6}, draw, top color=white, rounded corners=1pt,  minimum height=1.2cm, text width=1.8cm, align=center, text centered},
}


\tikzstyle{block} = [draw, rectangle, minimum height=3em]

\tikzstyle{output} = [coordinate]

\begin{tikzpicture}[auto, node distance=5cm,>=latex']


    \node [bnode, minimum width=7em, align=center] (ccdm) {Distribution Matcher};
    \node [left of=ccdm, node distance=2.3cm] (start) {};
    \node [bnode, minimum width=7em, right of=ccdm, node distance=8.6cm, align=center] (inencoder) {Systematic \\ FEC Encoder};

    \node [right of= inencoder, node distance=6.5cm, align=center](blank1){};
    
    \node [rectangle, draw=black, minimum width=7em,right of= inencoder, node distance=6.5cm, align=center,minimum width=3.1cm, minimum height=3.8cm,label={[label distance=0.2cm]90:OOK Modulator}](modulator){};
    \node [bnode, draw=black, minimum width=7em, above of= blank1, node distance =1cm, align=center](modulator1){Mapper\\$\{0,\AS\}$};
    \node [bnode, draw=black,minimum width=7em, below of=blank1, node distance=1cm, align=center](modulator2){Mapper\\$\{0,\AU\}$ };

    \node [bnode, minimum width = 7em, right of=blank1, node distance=5.4cm, align=center](mux){MUX};

    \node [bnode, minimum width=7em,xshift=3cm, yshift=-3cm, at =(mux), align=center] (channel) {AWGN\\Channel};

    \node [bnode, minimum width=7em,xshift=-2cm, yshift=-3cm, at =(channel), align=center] (demux) {DEMUX};

 \node [rectangle, draw=black, minimum width=7em,xshift=-5cm, at =(demux), align=center,minimum width=3.1cm, minimum height=3.8cm,label={[label distance=0.2cm]90:OOK Demodulator}](modulator){};
    \node [bnode, minimum width=7em, xshift=-5cm, yshift=1cm, at =(demux), align=center] (demod1) {Demodulator\\$\{0,\AS\}$};
    \node [bnode, minimum width=7em, xshift=-5cm, yshift=-1cm, at =(demux), align=center] (demod2) {Demodulator\\$\{0,\AU\}$};

    \node [draw,circle, minimum width=1em,  xshift=-5cm, yshift=0cm, at =(demod1), align=center] (plus) {+};
    
        
        \coordinate [ xshift=-5cm, yshift=0cm, at =(demod2), align=center] (plus2) ;

        
        \node [draw = none, above of = plus, node distance = 1.3cm, align=center] (prior) {$ \log \left({\frac{P_{\XS}(\AS)}{P_{\XS}(0)  }}\right)$};

    \node [bnode, minimum width=7em, left of=demux, node distance=12.5cm, align=center] (mux2) {MUX};
    \node [bnode, minimum width=7em, left of=mux2, node distance=5.5cm, align=center] (outdecoder) {FEC Decoder};

    \node [bnode, minimum width=7em, left of=outdecoder, node distance=3.5cm, align=center] (dematcher) {Dematcher};
    \node [left of=dematcher, node distance=2.3cm] (end) {};    
        

    \draw [->] (start) -- (ccdm);
    \draw [->] (ccdm) -- node[label={[label distance=-20mm]-20:$\bm{u} = (u_0, u_1, \ldots, u_{k-1})$}] {} (inencoder);
    \draw [->] ($(inencoder.east)+(0,0.25)$) to[-|-] node[label={[label distance=5mm]-150:$\bm{u}$}] {} (modulator1);
    \draw [->] ($(inencoder.east) -(0,0.25)$) to[-|-] node[label={[label distance=-14mm]130:$\bm{p}= (p_0, p_1, \ldots, p_{n-k-1})$}] {}(modulator2);


    \draw [->] (modulator1) to[-|-] node[label={[label distance=4mm]30:$\bm{x}_{\mathsf{S}}$}] {} ($(mux.west)+(0,0.25)$);
    \draw [->] (modulator2) to[-|-] node[label={[label distance=4mm]-30:$\bm{x}_{\mathsf{U}}$}] {} ($(mux.west)-(0,0.25)$);

    \draw [->] (channel.south) |- node [name=y1] {$\bm{y}$}(demux.east);

    \draw [->] (mux) -|  node [label={[label distance=-15mm]-30:$\bm{x} = (\bm{x}_{\mathsf{S}} \hspace{1mm} \bm{x}_{\mathsf{U}}) $}] {}(channel.north);

    \draw [->] ($(demux.west)+(0,0.25)$) to[-|-] node[label={[label distance=4mm]-20:$\bm{y}_{\mathsf{S}}$}] {} (demod1);
    \draw [->] ($(demux.west) -(0,0.25)$) to[-|-] node[label={[label distance=4mm]30:$\bm{y}_{\mathsf{U}}$}] {}(demod2);




    \draw [->] (prior) --   (plus);

    \draw [->] (demod1.west) to[-|-] node[label={[label distance=-20mm]30:$\left( L_0^{\mathsf{ch}}, L_1^{\mathsf{ch}}, \ldots, L_{k-1}^{\mathsf{ch}} \right)$}] {} (plus);

    \draw [->] (demod2.west) to [-] node[label={[label distance=-25mm]20:$\left( L_k^{\mathsf{ch}}, L_{k+1}^{\mathsf{ch}}, \ldots, L_{n-1}^{\mathsf{ch}} \right)$}] {}  (plus2) to [-|-] ($(mux2.east) -(0, 0.24)$);

    \draw [->] (plus) to[-|-]  ($(mux2.east) +(0,0.24)$);
   
    \draw [->] (mux2) --  node[label={[label distance=-18mm]30:$\left( L_0, L_1, \ldots, L_{n-1} \right)$}] {}  (outdecoder);
    \draw [->] (outdecoder) -- node [name=w] {$\bm{\hat{u}}$} (dematcher);
    \draw [->] (dematcher) --   (end);
\end{tikzpicture}


%% file: figures/achievable_rates_ps-Rc=0.50.tex
\begin{tikzpicture}
\begin{axis}[
 width=0.5\textwidth,
 height=0.4\textwidth,
xmin=-6,
xmax=6,
xlabel={$E_s/N_0$ [\unit{dB}]}, 
ylabel={Rate [\unit{bits/channel use}]},
grid=major,
ymode=linear,
legend cell align={left},
major grid style={dotted,gray},
legend style={at={(0.01,0.99)},anchor=north west,thick,font=\scriptsize },
legend pos=south east,
]

\addplot[name path global=ook_capacity,thick, black,solid,mark=.]
table[x=Es/N0,y=se2] {./data/cap_uni_opt.txt};
\addlegendentry{OOK capacity};
\addplot[name path global=ook_uni,thick,red,solid,mark=.]
table[x=Es/N0,y=se1] {./data/cap_uni_opt.txt};
\addlegendentry{OOK uniform};

\addplot[name path global=rts1,thick,TUMBeamerBlue,solid,mark=.]
table[x=Es/N0,y=rts1] {./data/achievable_rates_ps-Rc=0.50.txt};
\addlegendentry{$\RTS^*_1$};
\addplot[name path global=rtx1,thick,TUMBeamerBlue,dashed]
table[x=Es/N0,y=rtx1] {./data/achievable_rates_ps-Rc=0.50.txt};
\addlegendentry{$\RTX$ Case 1};

\addplot[name path global=rts2,thick,TUMBeamerGreen,solid]
table[x=Es/N0,y=rts2] {./data/achievable_rates_ps-Rc=0.50.txt};
\addlegendentry{$\RTS_2^*$};
\addplot[name path global=rtx2,thick,TUMBeamerGreen,dashed]
table[x=Es/N0,y=rtx2] {./data/achievable_rates_ps-Rc=0.50.txt};
\addlegendentry{$\RTX$ Case 2};

\path[name intersections={of=rts1 and rtx1, name=p1}, name intersections={of=rts2 and rtx2, name=p2}];


\node[circle,fill,inner sep=1.5pt,TUMBeamerBlue] at (p1-1) {};
\node[circle,fill,inner sep=1.5pt,TUMBeamerGreen] at (p2-1) {};

\path[name path global=line1] (p1-1) -- (p1-1-|{rel axis cs:\pgfkeysvalueof{/pgfplots/xmin},0});

\path[name path global=line2] (p2-1) -- (p2-1-|{rel axis cs:\pgfkeysvalueof{/pgfplots/xmin},0});

\path[name intersections={of=line1 and ook_capacity, name=p3}, name intersections={of=line2 and ook_capacity, name=p4}];

\draw[<->,thick] let \p1=(p3-1), \p2=(p1-1) in (p3-1) -- (p1-1) node [below,xshift=-1.2cm,yshift=.3cm] {%
        \pgfplotsconvertunittocoordinate{x}{\x1}%
        \pgfplotscoordmath{x}{datascaletrafo inverse to fixed}{\pgfmathresult}%
        \edef\valueA{\pgfmathresult}%
        \pgfplotsconvertunittocoordinate{x}{\x2}%
        \pgfplotscoordmath{x}{datascaletrafo inverse to fixed}{\pgfmathresult}%
        \pgfmathparse{\pgfmathresult - \valueA}%
        \pgfmathprintnumber{\pgfmathresult} dB
};

\draw[<->,thick] let \p1=(p4-1), \p2=(p2-1) in (p4-1) -- (p2-1) node [below,xshift=-1.2cm,yshift=.25cm] {%
        \pgfplotsconvertunittocoordinate{x}{\x1}%
        \pgfplotscoordmath{x}{datascaletrafo inverse to fixed}{\pgfmathresult}%
        \edef\valueA{\pgfmathresult}%
        \pgfplotsconvertunittocoordinate{x}{\x2}%
        \pgfplotscoordmath{x}{datascaletrafo inverse to fixed}{\pgfmathresult}%
        \pgfmathparse{\pgfmathresult - \valueA}%
        \pgfmathprintnumber{\pgfmathresult} dB
};

\path[name path global=line3] (p1-1) -- (p1-1-|{rel axis cs:\pgfkeysvalueof{/pgfplots/xmax},0});

\path[name path global=line4] (p2-1) -- (p2-1-|{rel axis cs:\pgfkeysvalueof{/pgfplots/xmax},0});

\path[name intersections={of=line3 and ook_uni, name=p3}, name intersections={of=line4 and ook_uni, name=p4}];

\draw[<->,thick] let \p1=(p1-1), \p2=(p3-1) in (p1-1) -- (p3-1) node [below,xshift=0.8cm,yshift=-.1cm] {%
        \pgfplotsconvertunittocoordinate{x}{\x1}%
        \pgfplotscoordmath{x}{datascaletrafo inverse to fixed}{\pgfmathresult}%
        \edef\valueA{\pgfmathresult}%
        \pgfplotsconvertunittocoordinate{x}{\x2}%
        \pgfplotscoordmath{x}{datascaletrafo inverse to fixed}{\pgfmathresult}%
        \pgfmathparse{\pgfmathresult - \valueA}%
        \pgfmathprintnumber{\pgfmathresult} dB
};

\draw[<->,thick] let \p1=(p2-1), \p2=(p4-1) in (p2-1) -- (p4-1) node [below,xshift=0.4cm,yshift=-.1cm] {%
        \pgfplotsconvertunittocoordinate{x}{\x1}%
        \pgfplotscoordmath{x}{datascaletrafo inverse to fixed}{\pgfmathresult}%
        \edef\valueA{\pgfmathresult}%
        \pgfplotsconvertunittocoordinate{x}{\x2}%
        \pgfplotscoordmath{x}{datascaletrafo inverse to fixed}{\pgfmathresult}%
        \pgfmathparse{\pgfmathresult - \valueA}%
        \pgfmathprintnumber{\pgfmathresult} dB
};

\end{axis}
\end{tikzpicture}

%% file: figures/achievable_rates_ps-Rc=0.75.tex
\begin{tikzpicture}
\begin{axis}[
 width=0.5\textwidth,
 height=0.4\textwidth,
xmin=-2,
xmax=6,
xlabel={$E_s/N_0$ [\unit{dB}]}, 
ylabel={Rate [\unit{bits/channel use}]},
grid=major,
ymode=linear,
legend cell align={left},
major grid style={dotted,gray},
legend style={at={(0.01,0.99)},anchor=north west,thick,font=\scriptsize },
legend pos=south east,
]

\addplot[name path global=ook_capacity,thick, black,solid,mark=.]
table[x=Es/N0,y=se2] {./data/cap_uni_opt.txt};
\addlegendentry{OOK capacity};
\addplot[name path global=ook_uni,thick,red,solid,mark=.]
table[x=Es/N0,y=se1] {./data/cap_uni_opt.txt};
\addlegendentry{OOK uniform};

\addplot[name path global=rts1,thick,TUMBeamerBlue,solid,mark=.]
table[x=Es/N0,y=rts1] {./data/achievable_rates_ps-Rc=0.75.txt};
\addlegendentry{$\RTS_1^*$};
\addplot[name path global=rtx1,thick,TUMBeamerBlue,dashed]
table[x=Es/N0,y=rtx1] {./data/achievable_rates_ps-Rc=0.75.txt};
\addlegendentry{$\RTX$ Case 1};

\addplot[name path global=rts2,thick,TUMBeamerGreen,solid]
table[x=Es/N0,y=rts2] {./data/achievable_rates_ps-Rc=0.75.txt};
\addlegendentry{$\RTS_2^*$};
\addplot[name path global=rtx2,thick,TUMBeamerGreen,dashed]
table[x=Es/N0,y=rtx2] {./data/achievable_rates_ps-Rc=0.75.txt};
\addlegendentry{$\RTX$ Case 2};

\path[name intersections={of=rts1 and rtx1, name=p1}, name intersections={of=rts2 and rtx2, name=p2}];

\node[circle,fill,inner sep=1.5pt,TUMBeamerBlue] at (p1-1) {};
\node[circle,fill,inner sep=1.5pt,TUMBeamerGreen] at (p2-1) {};

\path[name path global=line1] (p1-1) -- (p1-1-|{rel axis cs:\pgfkeysvalueof{/pgfplots/xmin},0});

\path[name path global=line2] (p2-1) -- (p2-1-|{rel axis cs:\pgfkeysvalueof{/pgfplots/xmin},0});

\path[name path global=line3] (p1-1) -- (p1-1-|{rel axis cs:\pgfkeysvalueof{/pgfplots/xmax},0});

\path[name path global=line4] (p2-1) -- (p2-1-|{rel axis cs:\pgfkeysvalueof{/pgfplots/xmax},0});

\path[name intersections={of=line1 and ook_capacity, name=p3}, name intersections={of=line2 and ook_capacity, name=p4}];

\draw[<->,thick] let \p1=(p3-1), \p2=(p1-1) in (p3-1) -- (p1-1) node [below,xshift=-1.1cm,yshift=.3cm] {%
        \pgfplotsconvertunittocoordinate{x}{\x1}%
        \pgfplotscoordmath{x}{datascaletrafo inverse to fixed}{\pgfmathresult}%
        \edef\valueA{\pgfmathresult}%
        \pgfplotsconvertunittocoordinate{x}{\x2}%
        \pgfplotscoordmath{x}{datascaletrafo inverse to fixed}{\pgfmathresult}%
        \pgfmathparse{\pgfmathresult - \valueA}%
        \pgfmathprintnumber{\pgfmathresult} dB
};

\path[name intersections={of=line3 and ook_uni, name=p3}, name intersections={of=line4 and ook_uni, name=p4}];

\draw[<->,thick] let \p1=(p1-1), \p2=(p3-1) in (p1-1) -- (p3-1) node [below,xshift=.6cm,yshift=-.1cm] {%
        \pgfplotsconvertunittocoordinate{x}{\x1}%
        \pgfplotscoordmath{x}{datascaletrafo inverse to fixed}{\pgfmathresult}%
        \edef\valueA{\pgfmathresult}%
        \pgfplotsconvertunittocoordinate{x}{\x2}%
        \pgfplotscoordmath{x}{datascaletrafo inverse to fixed}{\pgfmathresult}%
        \pgfmathparse{\pgfmathresult - \valueA}%
        \pgfmathprintnumber{\pgfmathresult} dB
};

\end{axis}
\end{tikzpicture}

%% file: sections/design.tex
\section{Protograph-Based LDPC Code Design}\label{sec:design}

We now discuss the design of \ac{P-LDPC} codes~\cite{Thorpe2003} for the scheme discussed in Fig.~\ref{fig:concatenation}. Protographs are small bipartite graphs which serve as a  template for a larger \ac{LDPC} code \cite{Thorpe2003}. A protograph can be represented by an $M \times N$ {basematrix} $\matB$ which contains elements from $\mathds{N}_0$. An element $b_{i,j}$ indicates the number of parallel edges between a \ac{VN} $V_j$ and a \ac{CN} $C_i$. 

The \ac{LDPC} code is obtained by a copy and permute operations applied to the Tanner graph of the protograph.

We use P-EXIT analysis to determine the decoding threshold of a protograph ensemble~\cite{TenBrink1999, LC07}. The decoding threshold is the smallest $\SNR$ such that the probability of symbol error vanishes as the blocklength (as well as the number of decoding iterations) goes to infinity. To briefly describe the algorithm, during each decoding iteration the mutual information between a message at each \ac{VN}/\ac{CN} output and the corresponding codeword bit is tracked. The analysis assumes that the messages are Gaussian distributed and that they fulfill the consistency condition \cite{Richardson2001}. This implies that the mean $\mu_m$ and the variance $\sigma^2_m$ of the messages are related to each other as $\mu_m=\sigma^2_m/2$.

\subsection{Surrogate Channel Design}
For our setting, the decoder soft-information does not fulfill the consistency condition,
which is needed for the all-zero codeword assumption and the analysis by P-EXIT. Evaluating \eqref{eq:llr} for our \ac{AWGN} model, we obtain 

\begin{equation}
L=\underbrace{\frac{A}{\sigma^2} y - \frac{A^2}{2\sigma^2}}_\text{{channel \ac{LLR}}}  +\underbrace{\log \left( \frac{P_X(A)}{P_X(0)}\right)}_\text{{prior}}.
\end{equation}
The prior for the information symbols breaks the consistency condition. Observe that $y$ is a realization of a Gaussian \ac{RV} with mean $\mu \in \{ 0, A \}$ and variance $\sigma^2$. Thus, also the $L$-value is a realization of a Gaussian \ac{RV} with mean $\left(\pm \frac{A}{2 \sigma^2} + \log  \frac{P_X(A)}{P_X(0)} \right)$ and variance $\frac{\AS^2}{ \sigma^2}$.

As a workaround, we use a surrogate channel approach~\cite{Peng2006}, i.e., the code is evaluated and optimized for a channel which is different from the target one, but captures its characteristics.

Following \cite{Steiner2016}, we use an \ac{AWGN} channel with uniformly distributed inputs. We have $\tilde Y = \tilde X + \tilde N$ with $\tilde X \in \{0,A\}$ and $\tilde N \sim \mathcal{N}(0,\tilde\sigma^2)$ as a surrogate and establish equivalence between the surrogate and target channel by requiring
\begin{equation} \label{eq:surrogate}
\entr{\tilde X | \tilde Y} =  \entr{\XS|\YS} .
\end{equation} 

\subsection{EXIT Analysis for the Time Sharing Schemes}

In order to perform protograph \ac{EXIT} analysis we consider the following setup: the first $\RC N$ \acp{VN} are connected to a binary-input \ac{AWGN} surrogate channel with variance $\sigSsur$ as described previously.
The remaining $(1-\RC)N$ \acp{VN} are connected to a binary-input \ac{AWGN} channel with variance $\sigU$. The following modifications with respect to standard P-EXIT analysis are required:
\begin{enumerate}
\item Pick a target transmission rate $\RTX$ and determine the additional parameters (code rate $\RC$, pulse probability and amplitudes $\AS$ and $\AU$) as explained in Sec.~\ref{sec:signaling_fixed_transmission_rate}. The code rate guides the selection of the protograph dimensions $N$ and $M$.
\item For a $\SNR=1/(2\sigma^2)$, compute the corresponding noise variances $\sigS$ and $\sigU$ as
\begin{align} \label{eq:normalize_snr}
\sigS  =  \frac{\sigma^2}{\AS^2 \pp} \text{~~and~~}\sigU =  \frac{2\sigma^2}{\AU^2} .
\end{align}
\item For the \ac{AWGN} channel with variance $\sigS$, find a surrogate channel with conditional entropy fulfilling \eqref{eq:surrogate}. Denote the variance of this channel by $\sigSsur$.
\item Initialize the channel noise variances of the first $\RC N$ \acp{VN} of the protograph with $\sigSsur$ and of the last $(1-\RC) N$ \acp{VN} with $\sigU$.
\item For the target $\SNR$, determine the a posteriori mutual information at the protograph \acp{VN} (after a sufficiently large number of iterations) by standard protograph \ac{EXIT} analysis as described in \cite{LC07}.
\end{enumerate}
In order to obtain the iterative decoding threshold of a protograph code ensemble, the above procedure is repeated for different $\SNR$. The lowest $\SNR$, for which the a posteriori mutual information approaches one for all \acp{VN} is the iterative decoding threshold of the protograph ensemble.

\subsection{Protograph Search}

To find good protograph ensembles, we use differential evolution (DE)~\cite{Uch14}. DE is a genetic optimization algorithm that finds capacity approaching protograph ensembles for various settings. We allow for a maximum number of $M-1$ \acp{VN} of degree $2$ \cite{Divsalar2009} 
and set the highest base matrix entry to~$4$.

%% file: sections/results.tex
\section{Numerical Results}\label{sec:results}


\subsection{Asymptotic Results}
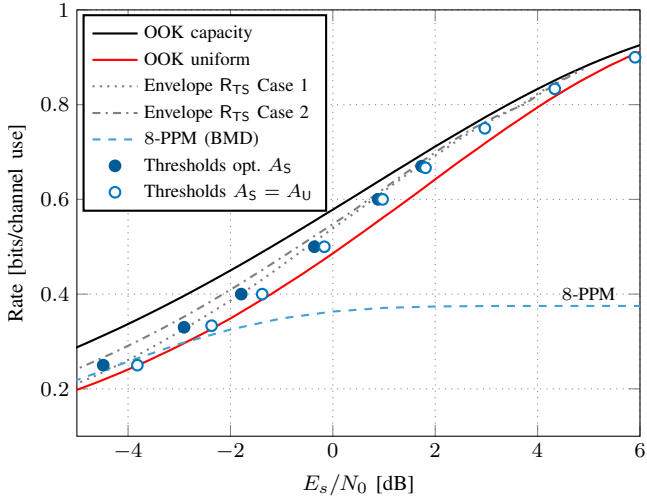
\begin{figure}[t]
    \centering
  \footnotesize
  \tikzsetnextfilename{rates_thresholds}
   \input{./figures/rates_thresholds}
   \caption{Achievable rates and thresholds versus $\SNR$ for various protographs.}
    \label{fig:asymptotic_results}
\end{figure}

We present the decoding thresholds for optimized protograph ensembles in Fig.~\ref{fig:asymptotic_results}. For the \ac{TS} scheme one, we consider $\RC \in \{0.5,0.67, 0.75, 0.8, 0.9\}$. For the \ac{TS} scheme two, we found optimized codes for $\RC \in \{0.67, 0.8\}$. A comparison to the achievable rates from Sec.~\ref{sec:signaling_fixed_transmission_rate} shows that the thresholds are close to the limits. 
At $\RTX=0.25$, we obtain a threshold of $\SI{-3.82}{dB}$ for \ac{TS} scheme one. For the \ac{TS} scheme two, the threshold is decreased to $\SI{-4.49}{dB}$. The gaps to the achievable rates are  $\SI{0.25}{dB}$ and  $\SI{0.3}{dB}$, respectively.

Note that for transmission rates $0.4 <\RTX < 0.85$ bits/channel use, \ac{TS} scheme one has a significant advantage with respect to uniform signaling and \ac{PPM} with \ac{BMD}. For  $\RTX<0.5$ bits/channel use, \ac{TS} scheme two gains with respect to \ac{PPM} with \ac{BMD}, uniform signaling and \ac{TS} scheme one. At $\RTX=0.25$, the protograph thresholds of \ac{TS} scheme two gain $\SI{0.67}{dB}$ of \ac{TS} scheme two over \ac{TS} scheme one.

\subsection{Simulation Results}

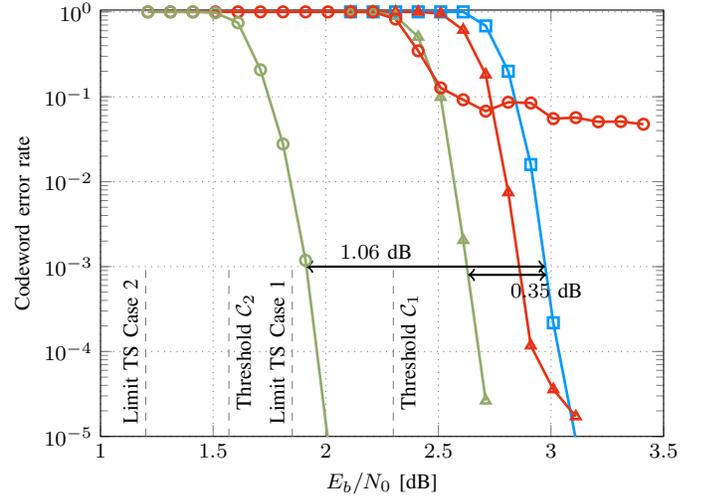
\begin{figure}[t]
    \centering
    \footnotesize
    \tikzsetnextfilename{coded_results_ook-Rtx=0.25}
     \input{./figures/coded_results_ook-Rtx=0.25}
    \tikzexternaldisable
    \caption{CER versus $E_b/N_0$ for $\RTX = \SI{0.25}{bpcu}$. The uniform reference (\ref{plt:uni0.25}) uses a DVB-S2 code of rate $\RC=\num{0.25}$. For TS Case~1, the optimized code (\ref{plt:ts1_opt_0.25}) and the DVB-S2 code (\ref{plt:ts1_dvbs2_0.25}) code have $\RC = \num{0.5}$. For TS Case~2 we have $\RC=0.67$ for the optimized (\ref{plt:ts2_opt_0.25}) and the DVB-S2 LDPC code~(\ref{plt:ts2_dvbs2_0.25}).}
    \tikzexternalenable
    \label{fig:CER_results_1}
\end{figure}

To verify our asymptotic findings, we construct finite length codes and compare to state-of-the-art off-the-shelf codes. Fig.~\ref{fig:CER_results_1} shows the \ac{CER} versus $E_b/N_0$ for two different \ac{P-LDPC} codes with $\RC=0.5$ ($\mathcal C_1$) and $\RC=0.67$ ($\mathcal C_2$) with a block length of $n=64800$ bits and for a transmission rate of $\RTX = \SI{0.25}{\text{bpcu}}$.
For $\RC=0.5$ we consider \ac{TS} scheme one while for $\RC=0.67$ we use  \ac{TS} scheme two.  
For comparison, the performance of an off-the-shelf DVB-S2 code~\cite{etsi2009dvb} with uniform signaling with $\RC=0.25$ is shown. Also, the performance of two off-the-shelf DVB-S2 codes with shaping (i.e., for $\RC=0.5$ and $\RC=0.67$) is shown.
We observe that in the waterfall region shaping gains \SI{0.1}{dB} for case 1 and \SI{0.35}{dB} for case 2, using codes from the DVB-S2 standard. However, the DVB-S2 \ac{LDPC} codes show visible error floors. Our designs gain \SI{0.35}{dB} for case 1 and \SI{1.1}{dB} for case 2, respectively. 

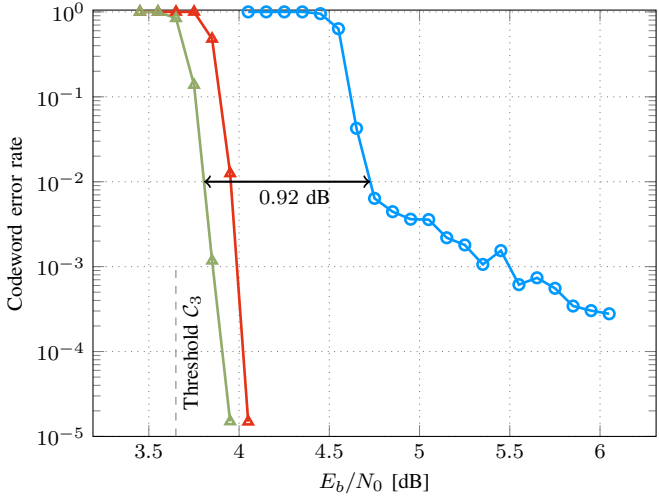
\begin{figure}[t]
    \centering
    \footnotesize
    \tikzsetnextfilename{coded_results_ook-Rtx=0.67}
   \input{./figures/coded_results_ook-Rtx=0.67}
   \tikzexternaldisable
    \caption{CER versus $E_b/N_0$ for $\RTX=\SI{0.67}{bpcu}$. The uniform reference (\ref{plt:uni0.67}) uses a DVB-S2 code of rate $\RC=\num{0.67}$. For TS Case~1, the optimized code (\ref{plt:ts1_opt_0.67}) and the DVB-S2 code (\ref{plt:ts1_dvbs2_0.67}) code have $\RC = \num{0.75}$.}
    \tikzexternalenable
    \label{fig:CER_results_2}
\end{figure}

Fig.~\ref{fig:CER_results_2} depicts the scenario for $\RTX =\SI{0.67}{\text{bpcu}}$ and \ac{TS} scheme one. Here we did not consider  \ac{TS} scheme two, since the achievable rate curves in Fig.~\ref{fig:asymptotic_results} suggest only small gains. Let $n=64800$ and $\RC = 0.75$ for $\mathcal C_3$. With shaping, the \ac{DVB-S2} code of $\RC = 0.75$ gains \SI{0.8}{dB} with respect to a \ac{DVB-S2} code of $\RC = 0.67$ with uniform signaling. A dedicated \ac{P-LDPC} code shows gains \SI{0.9}{dB} with respect to the uniform case.

%% file: figures/rates_thresholds.tex
\begin{tikzpicture}
\begin{axis}[
 width=0.5\textwidth,
 height=0.4\textwidth,
xmin= -5,
xmax=6,
ymin = 0.1,
ymax = 1,
xlabel={$\SNR$ [\unit{dB}]}, 
ylabel={Rate [\unit{bits/channel use}]},
grid=major,
xtick = {-20, -18,..., 15},
ymode=linear,
legend cell align={left},
major grid style={dotted,gray},
legend style={at={(0.01,0.99)},anchor=north west,thick,font=\scriptsize },
]

\addplot[line width=1, black,solid,mark=.,thick]
table[x=Es/N0,y=se2] {data/cap_uni_opt.txt};
\addlegendentry{OOK capacity};

\addplot[line width=1,red,solid,mark=.,thick]
table[x=Es/N0,y=se1] {data/cap_uni_opt.txt};
\addlegendentry{OOK uniform};

\addplot[line width=1,red,solid,mark=., gray,thick,dotted]
table[x=EsN0,y=C,col sep=space,trim cells=true] {data/envelope_rates_1p.txt};
\addlegendentry{Envelope $\RTS$ Case 1};

\addplot[line width=1,red,solid,mark=., gray,thick,dashdotted]
table[x=EsN0,y=C,col sep=space,trim cells=true] {data/envelope_rates_2p.txt};
\addlegendentry{Envelope $\RTS$ Case 2};

\addplot[line width=1,red,solid,mark=., TUMLighterBlue,thick,dashed]
table[x=SNR,y=C,col sep=space,trim cells=true] {data/capacity_Hadamard8_BICM.txt};
\addlegendentry{$8$-PPM (BMD)};

\addplot[only marks,thick, TUMDarkerBlue,solid,mark=*,mark options={fill=TUMDarkerBlue}]
table[x =esn0, y= rate] {data/best_thresholds_2p.txt};
\addlegendentry{Thresholds opt.\ $\AS$};

\addplot[only marks,thick, TUMMediumBlue,solid,mark=*,mark options={fill=white}]
table[x =esn0, y= rate] {data/best_thresholds_1p.tex};
\addlegendentry{Thresholds $\AS = \AU$};

 \node[] at (axis cs: 5,0.4) {\scriptsize $8$-PPM};

\pgfplotstableset{
    create on use/X/.style={create col/copy column from table={data/cap_uni_opt.txt}{Es/N0}}
}

\end{axis}
\end{tikzpicture}


%% file: figures/coded_results_ook-Rtx=0.25.tex
\begin{tikzpicture}

\begin{axis}[
 width=0.5\textwidth,
 height=0.4\textwidth,
xmin=1.0,
xmax=3.5,
ymin = 1e-5,
ymax = 1.05,
xlabel={$E_b/N_0$ [\unit{dB}]}, 
ylabel={Codeword error rate},
grid=major,
ymode=log,
legend cell align={left},
major grid style={dotted,gray},
legend style={at={(0.99,0.01)},anchor=south east,thick,font=\scriptsize },
]

\addplot[name path global=uni,line width=1,TUMBeamerBlue,mark=square,x filter/.code=\pgfmathparse{\pgfmathresult+3.0103}]
table[x = snr, y= fer] {./data/results-dvbs2-uni-Rtx=0.25.txt};\label{plt:uni0.25}

\addplot[line width=1,TUMBeamerRed,mark=triangle,x filter/.code=\pgfmathparse{\pgfmathresult+3.0103}]
table[x = snr, y= fer] {./data/results-ook-ts1-dvbs2-Rtx=0.25.txt};\label{plt:ts1_dvbs2_0.25}

\addplot[name path global=ts1,line width=1,TUMBeamerGreen,mark=triangle,x filter/.code=\pgfmathparse{\pgfmathresult+3.0103}]
table[x = snr, y= fer] {./data/results-ook-ts1-Rtx=0.25.txt};\label{plt:ts1_opt_0.25}

\addplot[line width=1,TUMBeamerRed,mark=o,x filter/.code=\pgfmathparse{\pgfmathresult+3.0103}]
table[x = snr, y= fer] {./data/results-ook-ts2-dvbs2-Rtx=0.25.txt};\label{plt:ts2_dvbs2_0.25}

\addplot[name path global=ts2,line width=1,TUMBeamerGreen,mark=o,x filter/.code=\pgfmathparse{\pgfmathresult+3.0103}]
table[x = snr, y= fer] {./data/results-ook-ts2-Rtx=0.25-code3.txt};\label{plt:ts2_opt_0.25}

\path[name path global=line] (1,1e-3) -- (3.5,1e-3);
\path[name path global=line1] (1,8e-4) -- (3.5,8e-4);

\path[name intersections={of=line and uni, name=p1}, name intersections={of=line and ts2, name=p3}];

\draw[<->,thick] let \p1=(p3-1), \p2=(p1-1) in (p3-1) -- (p1-1) node [above,xshift=-2.25cm] {%
        \pgfplotsconvertunittocoordinate{x}{\x1}%
        \pgfplotscoordmath{x}{datascaletrafo inverse to fixed}{\pgfmathresult}%
        \edef\valueA{\pgfmathresult}%
        \pgfplotsconvertunittocoordinate{x}{\x2}%
        \pgfplotscoordmath{x}{datascaletrafo inverse to fixed}{\pgfmathresult}%
        \pgfmathparse{\pgfmathresult - \valueA}%
        \pgfmathprintnumber{\pgfmathresult} dB
};

\path[name intersections={of=line1 and uni, name=p1}, name intersections={of=line1 and ts1, name=p2}];

\draw[<->,thick] let \p1=(p2-1), \p2=(p1-1) in (p2-1) -- (p1-1) node [below] {%
        \pgfplotsconvertunittocoordinate{x}{\x1}%
        \pgfplotscoordmath{x}{datascaletrafo inverse to fixed}{\pgfmathresult}%
        \edef\valueA{\pgfmathresult}%
        \pgfplotsconvertunittocoordinate{x}{\x2}%
        \pgfplotscoordmath{x}{datascaletrafo inverse to fixed}{\pgfmathresult}%
        \pgfmathparse{\pgfmathresult - \valueA}%
        \pgfmathprintnumber{\pgfmathresult} dB
};

\draw[dashed,gray] (axis cs: 2.3,1e-5) -- (axis cs: 2.3,1e-3) node[sloped,midway,below,black] {Threshold $\mathcal{C}_1$};
\draw[dashed,gray] (axis cs: 1.57,1e-5) -- (axis cs: 1.57,1e-3) node[sloped,midway,below,black] {Threshold $\mathcal{C}_2$};

\draw[dashed,gray] (axis cs: 1.8512,1e-5) -- (axis cs: 1.8512,1e-3) node[sloped,midway,above,black] {Limit TS Case 1};
\draw[dashed,gray] (axis cs: 1.2009,1e-5) -- (axis cs: 1.2009,1e-3) node[sloped,midway,above,black] {Limit TS Case 2};

\end{axis}
\end{tikzpicture}

%% file: figures/coded_results_ook-Rtx=0.67.tex
\begin{tikzpicture}

\begin{axis}[
 width=0.5\textwidth,
 height=0.4\textwidth,
ymin = 1e-5,
ymax = 1.05,
xlabel={$E_b/N_0$ [\unit{dB}]}, 
ylabel={Codeword error rate},
grid=major,
ymode=log,
legend cell align={left},
major grid style={dotted,gray},
legend style={at={(0.99,0.01)},anchor=south east,thick,font=\scriptsize },
]

\addplot[name path global=uni,line width=1,TUMBeamerBlue,mark=o,x filter/.code=\pgfmathparse{\pgfmathresult-1.2494}]
table[x = snr, y= fer] {./data/results-dvbs2-uni-Rtx=0.67.txt};\label{plt:uni0.67}

\addplot[line width=1,TUMBeamerRed,mark=triangle,x filter/.code=\pgfmathparse{\pgfmathresult-1.2494}]
table[x = snr, y= fer] {./data/results-ook-ts1-dvbs2-Rtx=0.67.txt};\label{plt:ts1_dvbs2_0.67}


\addplot[name path global=ts1,line width=1,TUMBeamerGreen,mark=triangle,x filter/.code=\pgfmathparse{\pgfmathresult-1.2494}]
table[x = snr, y= fer] {./data/results-ook-ts1-Rtx=0.67-code2.txt};\label{plt:ts1_opt_0.67}

\draw[dashed,gray] (axis cs: 3.65,1e-5) -- (axis cs: 3.65,1e-3) node[sloped,midway,below,black] {Threshold $\mathcal{C}_3$};

\path[name path global=line] (3,1e-2) -- (6,1e-2);

\path[name intersections={of=line and uni, name=p1}, name intersections={of=line and ts1, name=p2}];

\draw[<->,thick] let \p1=(p2-1), \p2=(p1-1) in (p2-1) -- (p1-1) node [below,xshift=-1cm] {%
        \pgfplotsconvertunittocoordinate{x}{\x1}%
        \pgfplotscoordmath{x}{datascaletrafo inverse to fixed}{\pgfmathresult}%
        \edef\valueA{\pgfmathresult}%
        \pgfplotsconvertunittocoordinate{x}{\x2}%
        \pgfplotscoordmath{x}{datascaletrafo inverse to fixed}{\pgfmathresult}%
        \pgfmathparse{\pgfmathresult - \valueA}%
        \pgfmathprintnumber{\pgfmathresult} dB
};

\end{axis}
\end{tikzpicture}

%% file: sections/conclusions.tex
\section{Conclusions}\label{sec:conclusions}
We proposed a \ac{PS} technique for \ac{OOK}  modulated \ac{AWGN} channels. We design \ac{P-LDPC} codes using a surrogate \ac{AWGN} channel approach. The proposed \ac{PS} scheme outperforms standard \ac{OOK} with a uniform distribution by \SI{0.7}{dB} at a transmission rate of \SI{0.25}{\bpcu} if the parity and information \ac{OOK} symbols are constrained to have the same pulse amplitude. Different amplitudes gain \SI{1.1}{dB} at a transmission rate of \SI{0.25}{\bpcu}.

\section*{Acknowledgment}
The authors would like to thank Gerhard Kramer for his valuable comments on this work.

%% file: sections/appendix.tex
\section{Base Matrices of the Simulated Codes}\label{app:protos}

In the following, we provide the optimized base matrices $\vB_1, \vB_2$ and $\vB_3$ for code $\cC_1, \cC_2$ and $\cC_3$, respectively. For $\vB_1$ the first column is punctured.

\setcounter{MaxMatrixCols}{20}
\begin{align}
    \vB_1 &= \vect{&3 &0 &0 &1 &2 &0 &0 \\
&1 &0 &2 &0 &0 &1 &2 \\
&3 &0 &1 &2 &2 &1 &1 \\
&2 &1 &0 &0 &0 &0 &0}
 \qquad   \vB_2 = \vect{&1 &0 &1 &0 &2 &0 &0 &0 &3 \\
&4 &2 &3 &2 &4 &2 &1 &1 &3 \\
&3 &1 &4 &1 &1 &1 &2 &1 &1 }\\
    \vB_3 &= \vect{&4 &0 &1 &4 &0 &3 &0 &0 &2 &0 &0 &0\\
&0 &1 &2 &3 &1 &1 &1 &2 &2 &1 &3 &2 \\
&3 &2 &1 &4 &1 &2 &2 &1 &2 &1 &1 &1 } 
\end{align}